\documentclass[onecolumn,showpacs]{revtex4}
\usepackage{amssymb,amsmath}
\usepackage{graphicx}

\newcommand{\hk}{\hat{\mathbf{k}}}

\newcommand{\bn}{\hat{\mathbf{n}}}

\newcommand{\bk}{\mathbf{k}}

\newcommand{\bm}{\hat{\mathbf{m}}}

\newcommand{\beq}{\begin{equation}}
\newcommand{\eeq}{\end{equation}}
\newcommand{\bea}{\begin{eqnarray}}
\newcommand{\eea}{\end{eqnarray}}
\newcommand{\ba}{\begin{array}}
\newcommand{\ea}{\end{array}}
\newcommand{\rad}{r}    

\newlength{\sizeonefig}
\newlength{\sizetwofig}
\newcommand{\mnras}{Mon. Not. R. Astr. Soc.}

\newcommand{\Ylm}[1]{Y_{l_#1}^{m_#1}}

\newcommand{\Ylmn}{Y_{l}^{m}}

\newcommand{\alm}[1]{a_{l_#1 m_#1}}

\newcommand{\bean}{\begin{eqnarray*}}
\newcommand{\eean}{\end{eqnarray*}}

\newcommand{\bx}{{\bf x}}

\begin{document}

\title{The trispectrum of 21-cm background anisotropies  as a probe of primordial non-Gaussianity}

\author{Asantha Cooray$^1$, Chao Li$^2$, Alessandro Melchiorri$^3$}
\affiliation{
$^1$Center for Cosmology, Department of Physics and Astronomy, 4129  Frederick
Reines Hall, University of California, Irvine, CA 92697\\
$^2$Theoretical Astrophysics, California Institute
of Technology, Mail Code 103-33 Pasadena, California 91125\\
$^3$Dipartimento di Fisica ``G. Marconi'' and INFN, sezione  di Roma, Universita' di Roma ``La Sapienza'', Ple Aldo Moro 5,  00185, Roma, Italy.
}

\date{\today}

\begin{abstract}
The 21-cm anisotropies from the neutral hydrogen distribution prior to the era of reionization is a
sensitive probe of primordial non-Gaussianity. Unlike the case with cosmic microwave background,
21-cm anisotropies provide multi-redshift information with frequency selection and is not damped at arcminute angular scales.
We discuss the angular trispectrum of the 21-cm background anisotropies and discuss how the trispectrum
signal generated by the primordial non-Gaussianity can be measured with  the three-to-one 
correlator and the corresponding angular power spectrum. We also discuss the
separation of primordial non-Gaussian information in the trispectrum
with that generated by the subsequent non-linear gravitational evolution of the density field.
While with the angular bispectrum of 21-cm anisotropies one can limit the second order corrections 
to the primordial fluctuations  below $f_{\rm NL}\sim 1$,
using the trispectrum information we suggest that the third order  coupling term, $f_2$ or $g_{\rm NL}$,
can be constrained to be arounde 10 with future 21-cm observations over the redshift interval of 50 to 100.
\end{abstract}
\pacs{98.80.Es,95.85.Nv,98.35.Ce,98.70.Vc}

\maketitle

\section{Introduction}

The cosmic 21-cm background involving spin-flip line emission or absorption
of neutral hydrogen contains unique signatures  on how the neutral gas evolved from last
scattering at $z \sim 1100$ to complete reionization at $z < 10$ \cite{Furlanetto}.
Subsequent to recombination, the temperature of neutral gas is coupled to that of the cosmic microwave background (CMB). At redshifts below
$\sim$ 200 the gas cools adiabatically, its temperature drops below that of the CMB, and neutral hydrogen
resonantly absorbs CMB flux through the spin-flip transition \cite{field,loeb,Bharadwaj}. The inhomogeneous
neutral hydrogen density distribution generates anisotropies in the brightness temperature measured
relative to the blackbody CMB \cite{zaldarriaga}. The large cosmological and astrophysical
information content in 21-cm background is well understood in the
literature \cite{loeb,Iliev,Santos,Bowman,McQuinn,Pen,sigurdson}.

Parallel to the large effort in analytical and numerical calculations of the 21-cm properties,
there are now several first generation 21-cm experiments underway focusing on the 21-cm signal during the era of reionization.
At the low redshifts probed by these first generation interferometers, the 21-cm signal is modified by the  astrophysics of
first  sources and the associated UV photon background. There, one naturally expects fluctuations to be dominated by inhomogeneities in source properties \cite{Zahn,Santos2}.
The associated non-Gaussianity in 21-cm anisotropies leads to a measurable three-point correlation function or
a bispectrum \cite{Coo05,Bharadwaj2,morales2}. Such a non-Gaussianity is also expected to dominate the
signature in 21-cm anisotropies generated by the primordial non-Gaussian density field.
However, the 21-cm background generated by neutral hydrogen at redshifts of 50 to 100
prior to the onset of reionization and the appearance of first stars
is expected to provide a cleaner probe of the primordial density perturbations
in the same manner CMB observations are used to study primordial fluctuations \cite{loeb}.
If the primordial fluctuations are non-Gaussian, then the 21-cm anisotropies at these high redshifts
will naturally contain a signature associated with that non-Gaussianity \cite{Cooray,pillepich}

The primordial non-Gaussianity in the density field can be studied with the three-point and higher-order correlation
functions of the 21-cm background. In particular,
the second order corrections to the density perturbations generated by primordial non-Gaussianity
lead to a bispectrum with a dependence on the second-order correction to the
curvature perturbations, $f_{\rm NL}$  \cite{Komatsu,Bartolo3}.
With future low frequency data out to $z \sim 100$,  21-cm background anisotropies
could potentially limit $f_{\rm NL} < 0.1$ \cite{Cooray}.
In comparison, the expected non-Gaussianity under standard inflation is of order $|n_s-1|$
and with the scalar spectral index $n_s \sim 0.98$ \cite{Spergel}, $f_{\rm NL}$ is expected to be
 well below unity \cite{Maldacena}.
The primordial non-Gaussianity parameter at the second order $f_{\rm NL}$, however, has a correction
associated with evolution of second and higher-order perturbations after inflation \cite{Bartolo}.
For standard slow-roll inflation then $f_{\rm NL} = -5/12(n_s-1) + 5/6 +3/10f(k)$ where the last term is
momentum dependent. In this case $f_{\rm NL}$ is at the level of a few tenths and could be as high as 1.
The ability for 21-cm anisotropies to probe $f_{\rm NL}$ as low as 0.1 is important since even a perfect CMB
experiment limited by cosmic variance alone can
only restrict $f_{\rm NL} > 3$ \cite{Komatsu,Babich,Smith} while there is no significant improvement when
using low redshift large-scale structure \cite{Scoc,Dalal}.

The possibility to make primordial non-Gaussianity measurements with the 21-cm background is important since
compared to most other probes of inflationary parameters, $f_{\rm NL}$ is one of the few parameters
for which we have limited number of probes sensitive to the low amplitude non-Gaussianity expected under
standard slow-roll inflation. While CMB as a probe of non-Gaussianity is well known,
when compared to CMB temperature and polarization anisotropies, 
21-cm background has two distinct advantages:  (1) the ability to probe multiple redshifts based on frequency selection
and (2) the lack of a damping tail in the 21-cm anisotropy  spectrum, unlike damping of CMB anisotropies at a multipole around 2000.

While the 21-cm background as a potentially interesting and a useful probe of $f_{\rm NL}$ is now known, different scenarios for
primordial fluctuations may produce a small $f_{\rm NL}$ but a large third-order correction.
In certain alternative models of inflation higher order terms may be significant even if the second-order term is small
and such scenarios include
 the new ekpyrotic cosmology \cite{Buchbinder} and, under certain conditions, the curvaton model \cite{Sasaki}.
Thus, beyond the non-Gaussianity at the three-point level with the bispectrum, it is also useful to
study the non-Gaussianities of the 21-cm background at the four point-level involving the trispectrum.

Here we show that the angular trispectrum of 21-cm anisotropies contains a measurable non-Gaussianity from primordial
fluctuations if the scale-independent cubic corrections to the gravitational potential captured with an
amplitude parameter $f_2$ (or $g_{\rm NL}$ as described in Ref.~\cite{Amico}) has a value of order $\sim 10$, even if
the scale-independent quadratic corrections to the gravitational potential captured by $f_{\rm NL}$ has a value around $\sim$ 1.
While $f_{\rm NL}$ is currently constrained with WMAP data \cite{Spergel}, there is no real constraint on this third order
non-Gaussianity parameter. There are, however, theoretical expectations:
for slow-roll inflation, a trispectrum of the form $T_4=1/2\tau_{\rm NL} [P(k_1)P(k_2)P(k_3) + ...]$ is expected for curvature perturbations
\cite{Seery} with an expectation value of $\tau \lesssim r/50$ where $r$ is the tensor-to-scalar ratio 
($r \lesssim 0.6$ is recent CMB data \cite{Spergel}).
In such a model, there is also a direct connection between $\tau_{\rm NL}$ and $f_{\rm NL}$
such that $\tau_{\rm NL}=(6f_{\rm NL}/5)^2$ (see the discussion in Ref.~\cite{Kogo:2006}
for connections between coupling terms of various models for primordial trispectrum). While such a relation is generally assumed when constraining
$\tau_{\rm NL}$ \cite{Lyth}, in future it may be that one can directly test the above relation between $f_{\rm NL}$ and $\tau_{\rm NL}$ with data.

To measure the non-Gaussianity at the four-point level, we
introduce the three-to-one correlator statistic, extending the two-to-one correlator of Ref.~\cite{Coo01} and applied to WMAP data 
in Ref.~\cite{Szapudi}.
We optimize the angular power spectrum of the 3-1 correlator to detect the primordial trispectrum by appropriately filtering 21-cm anisotropy data
to remove the non-Gaussian confusion  generated by the non-linear evolution of gravitational perturbations. To do this properly, one
requires prior knowledge on the configuration dependence  of the primordial 21-cm trispectrum, but its amplitude is a free variable to be determined
from the data.

This paper is organized as following: we first discuss the bispectrum in 21-cm anisotropies associated with
primordial perturbations resulting from quadratic corrections to the primordial potential.
We discuss ways to measure this bispectrum in the presence of other non-Gaussian signals and
determine the extent to which $f_{\rm NL}$ can be measured from 21-cm background data.
In the numerical calculations described later, we
take a fiducial flat-$\Lambda$CDM cosmological model with
$\Omega_b=0.0418$, $\Omega_m=0.24$, $h=0.73$, $\tau=0.092$,
$n_s=0.958$, and $A(k_0 = 0.05\ \mathrm{Mpc}^{-1})=2.3\times10^{-9}$.
This model is consistent with recent measurements from WMAP \cite{Spergel}.

\section{Calculational Method}
\label{calculation}

The 21-cm anisotropies are observed as a change in the intensity of the
CMB due to line emission or absorption at an observed frequency $\nu$:
\begin{equation}
T_b(\bn,\nu) =  \frac{T_S - T_{\rm CMB}}{1+z} \, \tau(\bn,\nu)
\label{eq:dtb}
\end{equation}
where $T_S$ is the spin temperature of the neutral gas, $z$ is the
redshift corresponding to the frequency of observation ($1+z=\nu_{21}/\nu$, with
$\nu_{21} = 1420$ MHz) and $T_{\rm CMB} = 2.73 (1+z) K$ is the CMB temperature at redshift $z$.
The optical depth, $\tau$, in the hyperfine transition \cite{field}, when
accounted for density and velocity perturbations of a patch in the neutral hydrogen distribution \cite{loeb,zaldarriaga,Bharadwaj}, is
\begin{eqnarray}
\tau & = & \frac{ 3 c^3 \hbar A_{10} \, \bar{n}_{\rm H}}{16 k_B \nu_{21}^2 \, T_S \, H(z) }
\left(1+\delta_H - \frac{1+z}{H(z)}\frac{\partial v}{\partial r}\right)  \, ,
\label{eq:tauigm}
\end{eqnarray}
where $A_{10}$ is the spontaneous emission coefficient for the transition ($2.85 \times 10^{-15}$ s$^{-1}$),
$n_{\rm HI}$ is the neutral hydrogen density,
$\delta_H=(n_{\rm H}-\bar{n}_{\rm H})/\bar{n}_{\rm H}$ is the inhomogeneity in the density,
$v$ is the peculiar velocity of the neutral gas distribution, $r$ is the comoving
radial distance, and $H(z)$ is the expansion rate at a redshift of $z$.  For simplicity, we have dropped the dependences
in location $\bn$. The fluctuations in the CMB brightness temperature is
\begin{equation}
\delta T_b=\bar{T}_b\left[\left(1-\frac{T_{\rm CMB}}{\bar{T}_S}\right)\left(\delta_H- \frac{1+z}{H(z)}\frac{\partial v}{\partial r}\right) + \frac{T_{\rm CMB}}{\bar{T}_S} S \delta_H\right]  \, ,
\label{eqn:deltab}
\end{equation}
where $S(z)$ describes the coupling between fluctuations in the spin temperature and the neutral density distribution \cite{Bharadwaj},
$\delta_H$ is fluctuations in the neutral density distribution, and $v(r)$ is the peculiar velocity.
In above $\bar{T}_b \approx 26.7 {\rm mK} \sqrt{1+z}$ \cite{zaldarriaga}.

We describe fluctuations in the 21-cm background in terms of multiple moments defined as $a_{lm}(\nu) =\int d\bn Y^*_{lm}(\bn) \delta T_b(\bn,\nu)$
where
\begin{eqnarray}
a_{l m}(\nu) &=& \int d\bn Y^*_{l m}(\bn)  \int d\rad W_\nu(\rad)
\bar{T}_b(\rad)\left[\left(1-\frac{T_{\rm CMB}}{\bar{T}_S}\right)\left(\delta_H- \frac{1+z}{H(z)}\frac{\partial v}{\partial r}\right) + \frac{T_{\rm CMB}}{\bar{T}_S} S \delta_H\
\right] \int \frac{d^3{\bf k}}{(2\pi)^3} \delta_H(\bk,\rad) e^{-i \bk \cdot \bn \rad} \nonumber \\
&=& 4\pi(-i)^l \int d\rad  W_\nu(\rad) \int \frac{d^3{\bf k}}{(2\pi)^3} f_k(\rad) M(k) D(\rad) \Phi^{\rm prim}(\bk) Y^*_{lm}(\hk) \, .
\label{eqn:alm}
\end{eqnarray}
In above,
\begin{equation}
f_k(\rad) = \bar{T}_b(\rad) \left[\left(1-\frac{T_{\rm CMB}}{\bar{T}_S}\right)J_l(k\rad) + \frac{T_{\rm CMB}}{\bar{T}_S}Sj_l(k\rad)\right] \, ,
\end{equation}
with $J_l(x)=j_l(x)-j_l''(x)$.
In equation~(\ref{eqn:alm}), $D(\rad)$ is the growth function of density perturbations, $M(k)=-3k^2T(k)/5\Omega_mH_0^2$ maps the
primordial potential fluctuations to that of the density field.

For simplicity in notation, hereafter, we will drop the explicit dependence on the frequency, but it should
be assumed that 21-cm background measurements can be constructed as a function of frequency,  and thus as a function of
redshift, with the width in frequency space primarily limited by the bandwidth of a radio interferometer captured by the window function $W_\nu(\rad)$.

Similar to calculations related to the CMB anisotropies, we can introduce a transfer function and rewrite multipole
 moments of 21-cm anisotropies as
\begin{equation}
a_{l m}= 4\pi(-i)^l \int \frac{d^3{\bf k}}{(2\pi)^3}\Phi^{\rm prim}(\bk) g_{\rm 21cm,l}(k)Y^*_{lm}(\hk) \, ,
\end{equation}
with the transfer function of the 21-cm background anisotropies given by
\begin{equation}
g_{\rm 21cm,l}(k) = \left[\int d\rad  W_\nu(\rad)  f_k(\rad) D(\rad)\right]M(k) \, .
\end{equation}
For reference, the angular power spectrum of 21-cm anisotropies
follows by taking  $\langle a_{lm}
a^*_{l'm'}\rangle=C_l\delta_{ll'}\delta_{mm'}$ as
\begin{equation}
C_l(\nu) = \frac{2}{\pi} \int k^2 dk P_{\Phi \Phi}(k) g^2_{\rm 21cm,l}(k)\, ,
\end{equation}
where the power spectrum of Newtonian potential is defined such that $\langle \Phi_L(\bk) \Phi_L^*(\bk')\rangle=(2\pi)^3 \delta_D(\bk-\bk')P_{\Phi \Phi}(k)$.

\subsection{21-cm Trispectrum}

Using the multiple moments of the 21-cm background $a_{lm}$,
 we can construct the angular average trispectrum, which is a rotationally invariant correlation function \cite{Hu:2001}.
We define this the usual way such that
\begin{equation}
\langle a_{l_1 m_1} a_{l_2 m_2} a_{l_3 m_3} a_{l_4 m_4} \rangle = \sum_{LM} (-1)^M
\left(
\begin{array}{ccc}
l_1 & l_2 & L \\
m_1 & m_2  &  -M
\end{array}
\right) \left(
\begin{array}{ccc}
l_3 & l_4 & L \\
m_3 & m_4  &  M
\end{array}
\right) T^{l_1 l_2}_{l_3 l_4}(L) \, ,
\end{equation}
where $T^{l_1 l_2}_{l_3 l_4}(L)$ is the angular averaged trispectrum. Here $l_1,l_2,l_3,l_4$ form a quadrilateral with $L$ as the length of the diagonal
and matrices are the Wigner 3-$j$ symbols. These symbols are non-zero under these conditions: $|l_1-l_2| \leq L \leq l_l+l_2$, $|l_3-l_4| \leq L \leq l_3+l_4$,
$l_1+l_2+L=$even, $l_3+l_4+L=$even, $m_1+m_2=M$, and $m_3+m_4=-M$.

The angular averaged trispectrum has two parts  involving a
connected piece associated with non-Gaussianity and an disconnected
part that is non-zero even if fluctuations are Gaussian with
${T_{\rm tot}}^{l_1 l_2}_{l_3 l_4}(L)={T_{\rm G}}^{l_1 l_2}_{l_3
l_4}(L)+{T_{\rm NG}}^{l_1 l_2}_{l_3 l_4}(L)$ \cite{Hu:2001}. The
Gaussian disconnected part is
\begin{eqnarray}
{T_{\rm G}}^{l_1 l_2}_{l_3 l_4}(L) &=& (-1)^{l_1+l_3} \sqrt{(2l_1+1)(2l_3+1)} C_{l_1} C_{l_3} \delta_{l_1 l_2} \delta_{l_3 l_4} \delta_{L0} \nonumber \\
&&\quad + (2L+1)C_{l_1}C_{l_2} \left[(-1)^{l_1+l_2+L}\delta_{l_1 l_3} \delta_{l_2 l_4} + \delta_{l_1 l_4} \delta_{l_2 l_3} \right] \, .
\end{eqnarray}
The connected part can be simplified based on permutation symmetries as
\begin{equation}
{T_{\rm NG}}^{l_1 l_2}_{l_3 l_4}(L)= P^{l_1 l_2}_{l_3 l_4}(L)+(2L+1)\sum_{L'}\left[(-1)^{l_2+l_3}\left\{
\begin{array}{ccc}
l_1 & l_2 & L \\
l_4 & l_3  &  L'
\end{array}
\right\}  P^{l_1 l_3}_{l_2 l_4}(L') + (-1)^{L+L'} \left\{
\begin{array}{ccc}
l_1 & l_2 & L \\
l_3 & l_4  &  L'
\end{array}
\right\} P^{l_1 l_4}_{l_3 l_2}(L') \right] \, ,
\end{equation}
where
\begin{equation}
P^{l_1 l_2}_{l_3 l_4}(L) = {{\cal T}_{\rm NG}}^{l_1 l_2}_{l_3 l_4}(L)+(-1)^{l_1+l_2+l_3+l_4+2L}{{\cal T}_{\rm NG}}^{l_2 l_1}_{l_4 l_3}(L)+(-1)^{l_3+l_4+L}{{\cal T}_{\rm NG}}^{l_1 l_2}_{l_4 l_3}(L)
+(-1)^{l_1+l_2+L}{\cal T}^{l_2 l_1}_{l_3 l_4}(L) \, .
\end{equation}
In above, matrices are now the Wigner 6$j$ symbol and ${\cal T}^{l_1 l_2}_{l_3 l_4}(L)$ is what is described as the reduced trispectrum in the literature \cite{Hu:2001,Kogo:2006,Okamoto:2002}.

To generate a non-Gaussianity in the 21-cm background that will lead to a trispectrum, we assume quadratic and cubic corrections to the
Newtonian curvature such that
\begin{equation}
\Phi(\bx)=\Phi_L(\bx)+f_{\rm NL} \left[\Phi^2_L(\bx)-\langle \Phi^2_L(\bx)\rangle\right] + f_2 \Phi^3_L(\bx) \,
\label{phi}
\end{equation}
when $\Phi_L(\bx)$ is the linear and Gaussian  perturbation and $f_{\rm NL}$ and $f_2$ are the coupling parameters, which may
or may not be scale dependent. Note that $f_2$ is also identified in some publications as $g_{\rm NL}$ \cite{Amico}
though we follow the notation of Ref.~\cite{Kogo:2006}.
Under this description, existing WMAP data limits $-30 < f_{\rm NL} < 74$
at the 1$\sigma$ confidence \cite{Szapudi},
while with an ideal CMB experiment fundamentally limited by cosmic variance  one can constrain $|f_{\rm NL}| < 3$ \cite{Komatsu}.
The angular bispectrum of 21-cm fluctuations, especially if measured between $z \sim 30$ and $z\sim 75$ prior to the
formation  of first sources and out to angular scales corresponding to multipoles of $\sim 10^4$ can be used to limit
$|f_{\rm NL}| < 0.1$ \cite{Cooray}.  This is well below the standard expectations for $f_{\rm NL}$
even after accounting for second-order evolution during horizon exit and re-entry \cite{Bartolo}. There are no useful
observational limits on $f_2$ and no estimate on how well $f_2$ can be established with the CMB trispectrum.

In Fourier space, we can decompose equation~(\ref{phi}) as
\begin{equation}
\Phi(\bk)=\Phi_L(\bk)+f_{\rm NL} \Phi_2(\bk) + f_2 \Phi_3(\bk) \, ,
\end{equation}
with
\begin{eqnarray}
\Phi_2(\bk) &=& \int \frac{d^3{\bk_1}}{(2\pi)^3}
\Phi_L(\bk+\bk_1) \Phi_L^*(\bk_1) - (2\pi)^3 \delta(\bk)
\int \frac{d^3{\bk_1}}{(2\pi)^3}P_{\Phi\Phi}(\bk_1) \\
\Phi_3(\bk) &=& \int \frac{d^3{\bk_1}}{(2\pi)^3} \int
\frac{d^3{\bk_2}}{(2\pi)^3}\Phi_L^*(\bk_1) \Phi_L^*(\bk_2)
\Phi_L(\bk_1+\bk_2+\bk)  \, .
\end{eqnarray}
Using these, the reduced connected trispectrum  can be written as
\begin{equation}
\langle \Phi(\bk_1)\Phi(\bk_2)\Phi(\bk_3)\Phi(\bk_4) \rangle_c = (2\pi)^3 \int d^3K \delta_D(\bk_1+\bk_2+{\bf K})\delta_D(\bk_3+\bk_4-{\bf K})
{\cal T}_\Phi(\bk_1,\bk_2,\bk_3,\bk_4;{\bf K})
\label{eqn:triphi}
\end{equation}
with two terms involving
\begin{eqnarray}
{\cal T}^{(2)}_\Phi(\bk_1,\bk_2,\bk_3,\bk_4;{\bf K}) &=& 4f_{\rm NL}^2P_{\Phi\Phi}(K)P_{\Phi\Phi}(k_1)P_{\Phi\Phi}(k_3) \nonumber \\
{\cal T}^{(3)}_\Phi(\bk_1,\bk_2,\bk_3,\bk_4;{\bf K}) &=& f_2\left[P_{\Phi\Phi}(k_2)P_{\Phi\Phi}(k_3)P_{\Phi\Phi}(k_4) + P_{\Phi\Phi}(k_2)P_{\Phi\Phi}(k_1)P_{\Phi\Phi}(k_4)\right]  \, .
\end{eqnarray}
Using the 21-cm transfer function defined in equation~(7) and using equation~(\ref{eqn:triphi}), we write
\begin{eqnarray}
&& \langle a_{l_1 m_1} a_{l_2 m_2} a_{l_3 m_3} a_{l_4 m_4} \rangle = (4\pi)^4 (-i)^{\sum l_i}
\int \frac{d^3{\bf k}_1}{\left( 2\pi \right) ^3}...\int\frac{d^3{\bf k}_4}{\left( 2\pi \right) ^3} \int d^3K \nonumber \\
&&\times (2\pi)^3 \delta_D(\bk_1+\bk_2+{\bf K})\delta_D(\bk_3+\bk_4-{\bf K}) {\cal T}_\Phi(\bk_1,\bk_2,\bk_3,\bk_4;{\bf K})
g_{\rm 21cm,l_1}(k_1) g_{\rm 21cm,l_2}(k_2) g_{\rm 21cm,l_3}(k_3) g_{\rm 21cm,l_4}(k_4) \nonumber \\
&&\times Y^*_{l_1m_1}(\hk_1) Y^*_{l_2m_2}(\hk_2) Y^*_{l_3m_3}(\hk_3) Y^*_{l_4m_4}(\hk_4)  \, .
\end{eqnarray}
We simplify further by expanding the $\delta_D$ functions, for example,
\begin{equation}
\delta_D(\bk_3+\bk_4-{\bf K}) = \int \frac{d^3{\bf r}}{(2\pi)^3} e^{-i {\bf r} \cdot (\bk_3+\bk_4-{\bf K})} \, ,
\end{equation}
and combining with the Rayleigh expansion of a plane wave
\begin{equation}
e^{i {\bf r} \cdot \bk} = (4\pi) \sum_{lm} i^l j_l(kr) Y^*_{lm}(\hk)Y^*_{lm}(\hat{\bf r})
\end{equation}
to simplify the reduced trispectrum of 21-cm anisotropies as
\begin{eqnarray}
&& {{\cal T}_{\rm NG}}^{l_1 l_2}_{l_3 l_4}(L) = \left(\frac{2}{\pi}\right)^5 \int k_1^2 dk_1...\int k_4^2 dk_1 \int K^2 dK \int r_1^2 dr_1 \int r_2^2 dr_2  {\cal T}_\Phi(\bk_1,\bk_2,\bk_3,\bk_4;{\bf K}) \\
&\times& g_{\rm 21cm,l_1}(k_1) g_{\rm 21cm,l_2}(k_2) g_{\rm 21cm,l_3}(k_3) g_{\rm 21cm,l_4}(k_4) j_{l_1}(k_1r_1) j_{l_2}(k_2r_1) j_{l_3}(k_3r_2) j_{l_4}(k_4r_2) j_{L}(Kr_1) j_{L}(Kr_2) h_{l_1LL_2} h_{l_3LL_4} \, , \nonumber
\label{eqn:tng}
\end{eqnarray}
where
\begin{equation}
h_{l_1L_2L}=\sqrt{\frac{(2l_1+1)(2l_2+1)(2L+1)}{4\pi}}
\left(
\begin{array}{ccc}
l_1 & l_2 & L \\
0 & 0  &  0
\end{array}
\right) \, .
\end{equation}

Substituting for the reduced trispectrum of Newtonian curvature, and similar to the CMB trispectrum \cite{Okamoto:2002,Kogo:2006},  the reduced trispectrum of 21-cm anisotropies can be written with two contributions from second- and third-order corrections as
\begin{eqnarray}
&& {{\cal T}_{\rm NG}}^{l_1 l_2}_{l_3 l_4}(L)  = 4f_{\rm NL}^2 \int r_1^2 dr_1 \int r_2^2 dr_2 F_L(r_1,r_2) \alpha_{l_1}(r_1) \beta_{l_2}(r_1) \alpha_{l_3}(r_2) \beta_{l_4}(r_2) h_{l_1l_2L}h_{l_3l_4L} \nonumber \\
&& \quad \quad + f_2 \int r^2 dr \beta_{l_2}(r) \beta_{l_4}(r)\left[\mu_{l_1}(r)\beta_{l_3}(r)+\beta_{l_1}(r)\mu_{l_3}(r)\right]h_{l_1 l_2 L}h_{l_3 l_4 L}
\label{eqn:prim}
\end{eqnarray}
where
\begin{eqnarray}
F_L(r_1,r_2) &=& \frac{2}{\pi} \int k^2 dk P_{\Phi \Phi}(k) j_L(kr_1)j_L(kr_2) \nonumber \\
\alpha_l(r) &=& \frac{2}{\pi} \int k^2 dk g_{\rm 21cm,l}(k)j_l(kr) \nonumber \\
\beta_l(r) &=& \frac{2}{\pi} \int k^2 dk P_{\Phi \Phi}(k) g_{\rm 21cm,l}(k)j_l(kr) \nonumber \\
\mu_l(r) &=& \frac{2}{\pi} \int k^2 dk g_{\rm 21cm,l}(k)j_l(kr)  \, .
\end{eqnarray}

\subsection{3-1 power spectrum}

Instead of measuring the full trispectrum to extract information on the non-Gaussianity of primordial perturbations captured by $f_{\rm NL}$ and $f_2$,
we consider a compact statistic that measures non-Gaussian information but can be described as a higher order 2-point statistic. For this
purpose, we introduce the three-to-one correlator analogous to the two-to-one correlator of Ref.~\cite{Coo01} and applied
to limit $f_{\rm NL}$ from WMAP in Ref.~\cite{Szapudi}. This statistic is
\begin{eqnarray}
W(\bn,\bm) &\equiv& \langle \hat{T}_b^3(\bn) T_b(\bm) \rangle  \\
           &=& \sum_{l_1 m_1 l_2 m_2} \langle \alm{1}^{3} \alm{2}^*
\rangle
               \Ylm{1}(\bn) \Ylm{2} {}^*(\bm)\, , \nonumber
\label{eqn:twopointsquared}
\end{eqnarray}
where $a_{lm}^3 = \int d\bn \hat{T}_b^3(\bn) \Ylmn {}^*(\bn)$, where $\hat{T}_b(\bn)$ is the 21-cm brightness temperature with the filter
described below applied in multipole space. We note that this statistic has been recently discussed in the context of separating lensing and kinetic
Sunyaev-Zel'dovich contributions to arcminute scale CMB anisotropies \cite{Riquelme}.

Similar to the filtered two-to-one angular power spectrum \cite{Coo01,Cooray},
the three-to-one cubic angular power spectrum is
\begin{equation}
\langle a_{lm}^3 a^*_{l'm'} \rangle = X_l^{\rm
tot}\delta_{ll'}\delta_{mm'} \, ,
\end{equation}
where
\begin{eqnarray}
X_l^{\rm tot} = \frac{1}{\sqrt{2l+1}} \sum_{l_1l_2l_3L} \frac{(-1)^{l_1+l_2+L}}{(2L+1)} {T_{\rm tot}}^{l_1l_2}_{l_3l}(L) w_{l_1l_2l_3|l,L}h_{l_1l_2L}h_{l_3lL} \, ,
\label{eqn:xl}
\end{eqnarray}
where $w_{l_1l_2l_3|l,L}$ is the form of the filter function that applies to the cubic field to optimize the detection of any particular
form of the underlying trispectrum. In real space, this filter can be simply described as $a_{lm}^3 = \int d\bn \hat{T}_b^3(\bn) W_1(\bn) W_2(\bn)W_3(\bn) \Ylmn {}^*(\bn)$ where $W_1$ to $W_3$ are real space filters that are applied to the three maps of the same field that are multiplied together.
In above $X_l$ note that contributions come from both the unconnected Gaussian part of the trispectrum ${T}_g$ and the connected non-Gaussian
part of the trispectrum $T_c$. For simplicity, we identify separately $X_l^{\rm prim}$ and $X_l^{\rm grav}$ as the contributions resulting from
the trispectra of the density field generated by primordial density perturbations  (equation~\ref{eqn:prim}) and the subsequent
non-linear gravitational  evolution of the density field (see below), respectively.
Here, we are primarily interested  in detecting the non-Gaussian information contained in the non-Gaussian part of the trispectrum produced
by primordial non-Gaussianities or $X_l^{\rm prim}$.  With the Gaussian contribution to the trispectrum, the
total three-to-one angular power spectrum can be written as $X_l^{\rm tot}=X_l^{\rm prim}+X_l^{\rm grav}+X_l^{\rm Gaussian}$.

Substituting Eq.~10 in in Eq.~19 we derive a simplified expression for $X_l^{\rm Gaussian}$ as
\begin{eqnarray}
X_l^{\rm Gaussian} = \frac{\sqrt{2l+1}}{4\pi}C_l\sum_{l_1L} (2l_1+1)C_{l_1} \left[w_{ll_1l|l,L}\delta_{L0} + (2L+1)
\left(
\begin{array}{ccc}
l_1 & l & L \\
0 & 0  &  0
\end{array}
\right)^2 \left\{w_{ll_1l|l,L}+w_{ll_1l_1|l,L}\right\}\right] \, .
\end{eqnarray}
If the filter function $w_{l_1l_2l_3|l,L}$ is designed such that it is equal to zero if any of $(l_1,l_2,l_3)$ is equal to another then
$X_l^{\rm Gaussian}=0$ and there is no contribution to the three-to-one angular power spectrum from the Gaussian term.

The variance of the three-to-one power spectrum calculated from $\langle X_l X_l' \rangle - \langle X_l\rangle^2$ with
\begin{equation}
\langle X_l X_l' \rangle = \frac{1}{(2l+1)(2l'+1)} \sum_{mm'}
\langle a_{lm}^3 a_{lm}^* a_{lm}^{3*} a_{lm}\rangle \, ,
\end{equation}
is
\begin{eqnarray}
(2l+1)N_l^2 &=& \left(X_l^{\rm prim}\right)^2 + \left(X_l^{\rm grav}\right)^2 +  \left(X_l^{\rm Gaussian}\right)^2 \nonumber \\
  && \quad \quad + \frac{C_l^{\rm tot}}{(2l+1)} \sum_{l_1l_2l_3L} \frac{w^2_{l_1l_2l_3|l,L}}{(2L+1)} C_{l_1}^{\rm tot}
C_{l_2}^{\rm tot} C_{l_3}^{\rm tot} h_{l_1l_2L}^2 h_{l_3lL}^2 \, ,
\end{eqnarray}
and we have ignored the covariance generated by non-Gaussian terms involving three- to eight-point correlations of $a_{lm}$.

To optimize the detection of $X_l^{\rm prim}$, under the assumption that $X_l^{\rm grav} < X_l^{\rm prim}$ first,
we find the shape of the filter that maximizes the signal-to-noise ratio for its detection is
\begin{equation}
w_{l_1l_2l_3|l,L} = \frac{(-1)^{l_1+l_2+L}}{h_{l_1l_2L}h_{l_3lL}}\frac{{T_c^{\rm prim}}^{l_1l_2}_{l_3l}}{C_{l_1}C_{l_2}C_{l_3}}\, ,
\label{eqn:filter}
\end{equation}
with the additional constraint that $w_{l_1l_2l_3|l,L}=0$ if two of $(l_1,l_2,l_3,l)$ are equal.

With the filter applied, the noise related to $X_l^{\rm prim}$ is
\begin{eqnarray}
N_l^{\rm tot} &=& \left[ \frac{\left(X_l^{\rm grav}\right)^2}{2l+1} +
\frac{C_l^{\rm tot}}{(2l+1)^2} \sum_{l_1l_2l_3L} \frac{w^2_{l_1l_2l_3|l,L}}{(2L+1)} C_{l_1}^{\rm tot}
C_{l_2}^{\rm tot} C_{l_3}^{\rm tot} h_{l_1l_2L}^2 h_{l_3lL}^2 \right]^{1/2} \, .
\label{eqn:nl}
\end{eqnarray}

With the assumption that $X_l^{\rm grav}\approx 0$ then maximum
signal-to-noise ratio for a detection of $X_l^{\rm prim}$ with a
noise spectrum of $N_l^{\rm tot}$ is
\begin{equation}
\left(\frac{\rm S}{\rm N}\right)^2 = \sum_{l_1>l_2>l_3>l,L} \frac{|{T_c^{\rm prim}}^{l_1l_2}_{l_3l}(L)|^2}{(2L+1)C^{\rm tot}_{l_1}C^{\rm tot}_{l_2}C^{\rm tot}_{l_3}C^{\rm tot}_{l}} \, .
\label{eqn:sntri}
\end{equation}
This is the same signal-to-noise ratio for a detection of the trispectrum generated by primordial non-Gaussianity.
However, $X_l^{\rm grav}$ is not necessarily zero and with the
same filter applied in the presence of non-negligible non-Gaussianity from non-linear density perturbations,
there is a residual contribution to $X_l$ that reduces the overall signal-to-noise ratio to be
below that of the maximal value for a detection of the primordial trispectrum alone. In this sense, in the presence of
secondary non-Gaussian signal, the filter in equation~(\ref{eqn:filter}) is non-optimal and could potentially be redesigned to
improve the overall signal-to-noise ratio for a detection of $X_l^{\rm prim}$. While we do not make such an attempt here,
in estimating the signal-to-noise ratio, we do account for the contamination from $X_l^{\rm grav}$ and the overall
signal-to-noise ratios we calculated from $X_l^{\rm prim}$ and $N_l$ is below the signal-to-noise ratio given in equation~(\ref{eqn:sntri}).

This degradation in the signal-to-noise ratio comes from the cross-correlation of trispectra of
primordial non-Gaussianity (as used in the filter) and the non-Gaussianity generated by gravitational evolution of density perturbations.
To understand this confusion associated
with non-linear gravitational evolution, we also calculate $X_l^{\rm grav}$ following the derivation
of ${T_c^{\rm grav}}^{l_1l_2}_{l_3l_4}(L)$  as
described in the Appendix. Since $w_{l_1l_2l_3|l,L}$ is defined in terms of the primordial trispectrum,
$X_l^{\rm grav} \propto \sum_{l_1l_2l_3L} {T_c^{\rm prim}}^{l_1l_2}_{l_3l}(L) \times {T_c^{\rm grav}}^{l_1l_2}_{l_3l}(L)$
while $X_l^{\rm prim} \propto \sum_{l_1l_2l_3L} |{T_c^{\rm prim}}^{l_1l_2}_{l_3l}(L)|^2$.
Since modes of ${T_c^{\rm grav}}^{l_1l_2}_{l_3l}(L)$ do not align with those of
${T_c^{\rm prim}}^{l_1l_2}_{l_3l}(L)$ the former
 sum has cancellations and in general we do expect $X_l^{\rm grav}$ to be at the same order as $X_l^{\rm prim}$ or below.

The dominant contribution to $N_l^{\rm tot}$ is not $X_l^{\rm grav}$  but is the Gaussian variance captured
by the second term of  equation~(\ref{eqn:nl}). This statement is independent of $f_{\rm NL}$ and $f_2$.
If stated differently, when properly filtered to search for the primordial non-Gaussianity,
the main confusion for detecting primordial signal is not the non-Gaussianity generated by
non-linear perturbations  but rather the Gaussian covariance associated with the statistical measurement of $X_l^{\rm prim}$.
In practice, the measurement of $X_l^{\rm prim}$ is likely to be further confused by the non-Gaussianity of
foregrounds, which we have mostly ignored in the present discussion.
Unfortunately, little is known about the expected level of the foreground intensity in the low radio frequency range of interest.
Techniques have been suggested and discussed to remove foregrounds below the detector noise levels \cite{Morales,Santos3}
and the filtering process we have outlined will further reduce the confusion from
the remaining residual foregrounds. This is clearly a topic for further study
once data become available with first-generation interferometers \footnote{http://www.lofar.org;
http://www.haystack.mit.edu/arrays/MWA}.

\begin{figure}[t]
\includegraphics[scale=0.5,angle=-90]{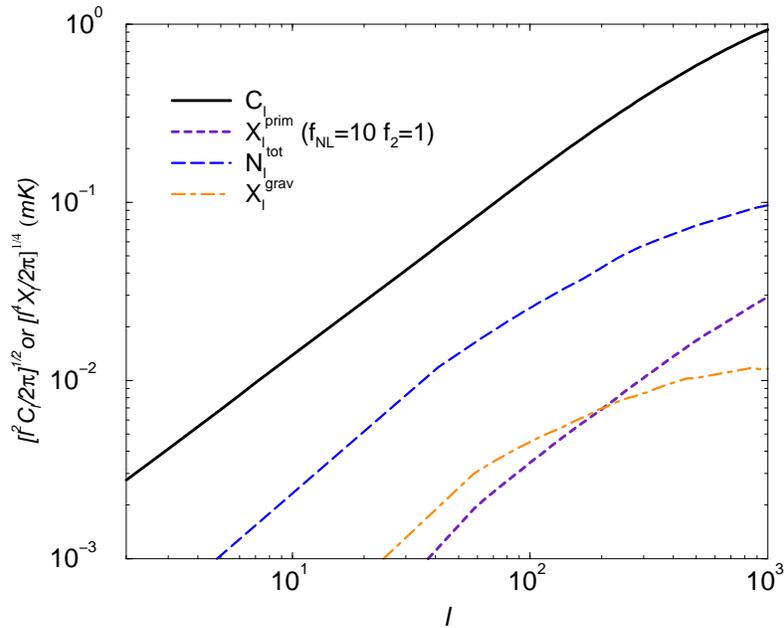}
\caption{
The power spectrum of 21-cm anisotropies (top solid line) and the angular power spectrum of the projected three-to-one correlator
described in the paper as a probe of the 21 cm trispectrum. In addition to $X_l^{\rm prim}$ with $f_{\rm NL}=10$ and $f_2=1$, we also show
$X_l^{\rm grav}$ and $N_l$.}
\label{spectra}
\end{figure}

\section{Results and Discussion}
\label{sec:results}

In Fig.~1 we summarize the power-spectrum of 21-cm anisotropies generated
by the neutral hydrogen distribution at a redshift of 100 with a bandwidth for observations of 1 MHz.
Here, we also plot $X_l^{\rm prim}$, $X_l^{\rm grav}$,
and $N_l^{\rm tot}$ for the same redshift with the optimal filter applied with $f_{\rm NL}=10$ and $f_2=1$ as
the non-Gaussian scale-independent amplitude of the primordial second- and third-order curvature perturbations, respectively.
As shown, $N_l^{\rm tot} > X_l^{\rm grav}$, suggesting that the noise term is dominated by the Gaussian variance (second term in equation~\ref{eqn:nl}).
This statement is independent of $f_{\rm NL}$ and $f_2$ and thus the non-Gaussian detection is dominated by the Gaussian
term in $N_l$ regardless of what is assumed about non-Gaussianity. Note that we have estimated $N_l^{\rm tot}$ in Fig.~1 under the
assumption that observations are limited only by the cosmic variance and not accounting for any instrumental noise variance,
which will also lead to a cut-off in $l$ out to which we can make measurements.
Using the cosmic variance alone allows us to establish the potentially achievable
limit and compare directly with cosmic variance limit with CMB data.
When calculating $X_l^{\rm prim}$ and $X_l^{\rm grav}$ in equation~(\ref{eqn:xl}) we set maximum value of $L$
in the sum associated with $T^{{l_1l_2}_{l_3l}}(L)$ to be $L_{\rm max}=100$
We tested our calculation for $L_{\rm max}=150$ and found results to be within a percent, but such a higher value slows the numerical calculation
significantly.  Finally, due to computational limitations of the numerical calculation, we restrict estimate of $X_l$ to $l=10^3$.

In Fig.~1 when calculating $X_l^{\rm grav}$, following the derivation in the Appendix and the
discussion there, we use the
exact analytical result for the 2-2 trispectrum of the density field with the mode coupling captured by $F_2(\bk_1,\bk_2)$ \cite{Goroff,fry84}.
For the 3-1 trispectrum of the density field under non-linear gravitational evolution, given that an
analytical result for the angular trispectrum is cumbersome, we use the angular averaged
value for $F_3(\bk_1,\bk_2,\bk_3)$. We refer the reader to the Appendix for details.

In Fig.~2 we summarize the estimate related to signal-to-noise ratio for a detection of $X_l^{\rm prim}$ as a function  of $l$.
The typical signal-to-noise ratio, when measurements are out to a multipole of $10^3$, is at the level of $\sim$ 0.5 if
one assumes that the coupling parameters $f_{\rm NL}=10$ and $f_2=3$ for 21-cm observations centered at a redshift of 100
over a bandwidth of 1 MHz.
The signal-to-noise ratio for the case with $f_{\rm NL}=0$ and $f_2=3$ is $\sim$ 0.03. Since $S/N \propto f_2$ when
$f_{\rm NL}=0$, out to $l=10^3$, a signal-to-noise ratio of 1 is achieved if $f_2 \sim 10^2$. While there are neither strong
theoretical motivations on the expected value of $f_2$ nor a real bound on its value from existing data, it is likely that with
21-cm data one can constrain $f_2$ down to a level well below this value for a single redshift. This is due to the fact that
21-cm observations lead to measurements at multiple redshifts, though
one cannot make arbitrarily small bandwidths to improve the detection since at scales below a
few Mpc, anisotropies in one redshift bin will be correlated with those in adjacent bins \cite{Santos3}. For example,
if 21-cm observations are separated to 30 independent bins over the redshift interval 50 to 100 (as can be achieved with 1 MHz bandwidths),
then an approximate estimate of the cumulative signal-to-noise ratio, ${\rm S}/{\rm N} = \sqrt{\sum_z [{\rm S}/{\rm N}(z)]^2}$,
is $\sim 0.15 (f_{2}/3)$ if $f_{\rm NL}=0$. In return, one can potentially probe $f_{\rm NL}$ values as low as 20 roughly
out to $l_{\rm max}\sim 10^3$.  With  the first-generation  radio interferometers, we would at most survey 1\% of the sky.
Assuming instrument noise is dominating at multipoles above $10^3$ between 30 MHz and 60 MHz at 1 MHz
bandwidths (corresponding to redshifts 30 to 100), we find a signal-to-noise ratio of $5 \times 10^{-3} f_{2}$, which could lead to
a limit on $f_2$ of order 200.

Above discussion on the application of $X_l^{\rm prim}$ assumes that $f_{\rm NL}=0$. Since $X_l^{\rm  prim} \propto f_{\rm NL}^2$,
if $f_{\rm NL}$ is greater than one, the overall signal-to-noise ratio for the detection of the three-to-one angular
power spectrum is increased and the dominant contribution to $X_l^{\rm prim}$ comes from the coupling
associated with $f_{\rm NL}$ and not $f_2$.  In the case when $f_2=0$, the $f_{\rm NL}$ one probes with the trispectrum can
be related to $\tau_{\rm NL}$ of Refs.~\cite{Seery,Lyth} for the primordial trispectrum. In Fig.~2, we show the signal-to-noise ratio with
$f_{\rm NL}=10$ and $f_2=3$. Since in this case $f_{\rm NL}$ term dominates, this provides an approximate estimate of the signal-to-noise
ratio for $f_{\rm NL}$ with the trispectrum. Using a single redshift bin out to $l_{\rm max} \sim 10^3$,
21-cm observations  achieve a signal-to-noise ratio of 1 if $f_{\rm NL} \sim 15$. This in return constraints $tau_{\rm NL} \lesssim 300$.
Assuming 30 redshift bins over the redshift interval 50 to 100, assuming $f_2=0$, we find that one can constrain $\tau_{\rm NL} < 50$.
This result is only out to $l_{\rm max} =10^3$, but since 21-cm observations are not damped as in the case of CMB observations,
higher resolution data can improve limits on both $f_2$ and $\tau_{\rm NL}$ significantly especially if observations can be
pushed to $l > 10^4$.

While a detection of the CMB angular trispectrum has been motivated as a way to constrain
$f_{\rm NL}$ or $\tau_{\rm NL}$ \cite{Kogo:2006}, this is probably not necessary with 21-cm data. Once radio interferometers start probing the
redshift interval of 50 to 100, the angular bispectrum, which can be probed with the two-to-one angular power spectrum \cite{Cooray},
can limit $f_{\rm NL} < 0.1$. This will facilitate a separation of the contribution to the three-to-one power spectrum from
$f_{\rm NL}$ and $f_2$. While we have assumed that $f_{\rm NL}$ and $f_2$ are momentum independent, it is likely that for
specific models of inflation, these coupling terms are momentum dependent \cite{Bartolo2}
and then the ability to separate $f_{\rm NL}$ and $f_2$ by combining 21-cm bispectrum and trispectrum information with the two-to-one
and three-to-one correlator respectively will strongly depend on the exact momentum dependence of the coupling factors.
We leave such a study for future research. While measuring $f_2$ is well motivated, one can also test the consistency between
$f_{\rm NL}$ probed by the bispectrum and the $\tau_{\rm NL}$ from the trispectrum related to the slow-roll inflation predictions for the
non-Gaussianity. While we have not discussed the extent to which this consistency relation can be established with upcoming
experiments after taking into account of instrumental noise it will also be useful to return to such  a calculation in the future once
21-cm observations begin to probe the universe at $z > 10$.

\begin{figure}[t]
\includegraphics[scale=0.5,angle=-90]{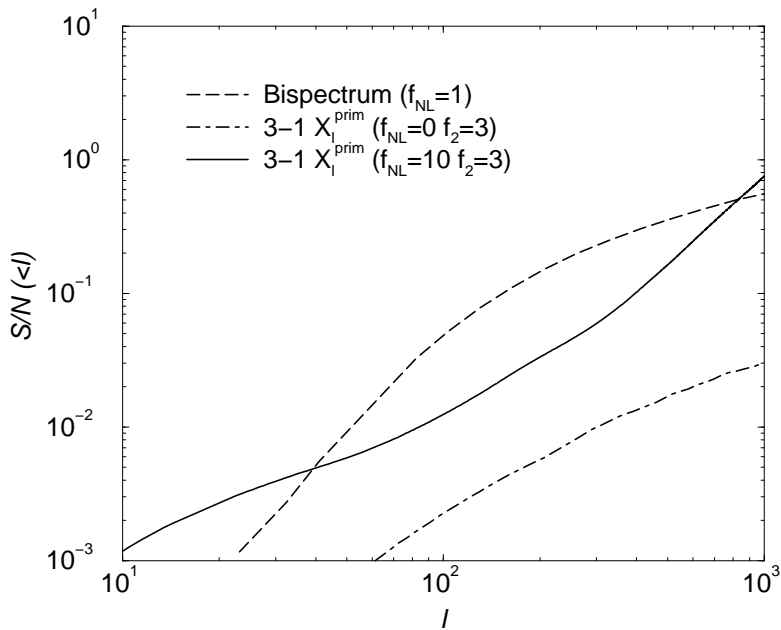}
\caption{
Signal-to-noise ratio for a detection of $X_l^{\rm prim}$ with $f_{\rm NL}=10$ and $f_2=3$ (solid line) and $f_{\rm NL}=0$ and $f_2=3$
(dot-dashed line). For reference, we show the signal-to-noise ratio associated with the detection of the full 21-cm bispectrum
with a dashed line.
} \label{sn}
\end{figure}

\section{Summary}
\label{sec:summary}

The 21-cm anisotropies from the neutral hydrogen distribution prior to the era of reionization is expected to
be more sensitive to primordial non-Gaussianity than the cosmic microwave background due to both
the three-dimensional nature of the 21-cm signal and the lack of a damping tail at arcminute angular scales.
Previous calculations have discussed the extent to which 21-cm bispectrum can be used as a probe of  primordial
non-Gaussianity at the two-point level with an non-Gaussianity parameter $f_{\rm NL}$ \cite{Cooray,pillepich}.

Here, we extend these calculations to discuss the possibility to use a four point statistic of the 21-cm background as a probe of the
primordial non-Gaussianity associated with the trispectrum.  We have calculated
 the angular trispectrum of the 21-cm background anisotropies and have introduced th
three-to-one correlator and the corresponding angular power spectrum as a probe of the primordial trispectrum captured by
both $f_{\rm NL}$ and the third order non-Gaussianity parameter $f_2$ (described in some publications as $g_{\rm NL}$).
Since the primordial non-Gaussianity is confused with a non-Gaussian signal in the 21-cm background generated by
the non-linear evolution of the density perturbations under gravitational evolution, we have discussed  way to
separate the two using an optimal filter.
While with the angular bispectrum of 21-cm anisotropies one can limit the second order corrections to the primordial fluctuations
as low as $f_{\rm NL}\sim 0.1$ below the value of $\sim 1$ expected for inflationary models, using the trispectrum information we
suggest that one can constrain third order coupling term $f_2$ to about few tens. If $f_{\rm NL}$ is large, it could potentially
be possible to test the consistency between $f_{\rm NL}$ from the bispectrum and the slow-roll non-Gaussianity $\tau_{\rm NL}$
at the four-point level with the relation $\tau_{\rm NL}=(6f_{\rm NL}/5)^2$. We hope to return to such a detailed study in the future.

\begin{acknowledgments}
This work was supported in part by NSF CAREER AST-0645427 at UC Irvine
and by the Moore Foundation at Caltech. We thank David Lyth for motivating us to calculate the 21 cm trispectrum.
\end{acknowledgments}


\section{Appendix}

Here we discuss the trispectrum from the non-linear density field.
The trispectrum is generated by both second and third-order perturbative corrections to the density fluctuations:
\begin{equation}
\delta(\bk)=\delta^{(1)}(\bk)+\delta^{(2)}(\bk)+\delta^{(3)}(\bk) \, ,
\end{equation}
where $\delta^{(1)}(\bk)=\delta_{\rm lin}(\bk)$ is the linear density perturbation and
\begin{eqnarray}
\delta^{(2)}(\bk) &=& \int \frac{d^3{\bf k}_1}{\left( 2\pi \right) ^3} \int \frac{d^3{\bf k}_2}{\left( 2\pi \right) ^3} (2\pi)^3 \delta_D(\bk_1+\bk_2-\bk) \delta_{\rm lin}(\bk_1)\delta_{\rm lin}(\bk_2) F_2(\bk_1,\bk_2) \\
\delta^{(3)}(\bk) &=& \int \frac{d^3{\bf k}_1}{\left( 2\pi \right) ^3} \int \frac{d^3{\bf k}_2}{\left( 2\pi \right) ^3} (2\pi)^3 \int \frac{d^3{\bf k}_3}{\left( 2\pi \right) ^3} (2\pi)^3 \delta_D(\bk_1+\bk_2+\bk_3-\bk) \delta_{\rm lin}(\bk_1)\delta_{\rm lin}(\bk_2) \delta_{\rm lin}(\bk_3)
F_3(\bk_1,\bk_2,\bk_3) \, , \nonumber
\end{eqnarray}
where $F_2(\bk_1,\bk_2)$ and $F_3(\bk_1,\bk_2,\bk_3)$ are derived in Ref.~\cite{Goroff,fry84}.

The reduced trispectrum of density perturbations $\langle \delta(\bk_1)\delta(\bk_2)\delta(\bk_3)\delta(\bk_4) \rangle$ can be written in terms of
the connected piece as
\begin{equation}
\langle \delta(\bk_1)\delta(\bk_2)\delta(\bk_3)\delta(\bk_4) \rangle = (2\pi)^3 \int d^3K \delta_D(\bk_1+\bk_2+{\bf K})\delta_D(\bk_3+\bk_4-{\bf K})
{\cal T}_\delta(\bk_1,\bk_2,\bk_3,\bk_4;{\bf K})
\label{eqn:tri}
\end{equation}
with two terms involving
\begin{eqnarray}
{\cal T}^{(2)}_\delta(\bk_1,\bk_2,\bk_3,\bk_4;{\bf K}) &=& 4F_2(\bk_1,{\bf K})F_2(\bk_3,{\bf K})P_\delta(K)P_\delta(k_1)P_\delta(k_3) \nonumber \\
{\cal T}^{(3)}_\delta(\bk_1,\bk_2,\bk_3,\bk_4;{\bf K}) &=& F_3(\bk_2,\bk_3,\bk_4)P_\delta(k_2)P_\delta(k_3)P_\delta(k_4) + F_3(\bk_2,\bk_1,\bk_4)P_\delta(k_2)P_\delta(k_1)P_\delta(k_4)  \, ,
\end{eqnarray}
where
\begin{eqnarray}
F_2(\bk_1,\bk_2) = \frac{5}{7}+\frac{\bk_1 \cdot \bk_2}{2k_2^2} +\frac{\bk_1 \cdot \bk_2}{2k_1^2} + \frac{2}{7} \left(\frac{\bk_1 \cdot \bk_2}{k_1k_2}\right)^2 \, ,
\end{eqnarray}
and $F_3(\bk_1,\bk_2,\bk_3)$ is derived in the Appendix of Ref.~\cite{Goroff}.

To calculate the angular trispectrum of 21-cm anisotropies we make use of the
 21-cm transfer function defined in equation~(7) and equation~(\ref{eqn:tri}) to write
\begin{eqnarray}
&& \langle a_{l_1 m_1} a_{l_2 m_2} a_{l_3 m_3} a_{l_4 m_4} \rangle = (4\pi)^4 (-i)^{\sum l_i}
\int \frac{d^3{\bf k}_1}{\left( 2\pi \right) ^3}...\int\frac{d^3{\bf k}_4}{\left( 2\pi \right) ^3} \int d^3K \nonumber \\
&&\times (2\pi)^3 \delta_D(\bk_1+\bk_2+{\bf K})\delta_D(\bk_3+\bk_4-{\bf K}) {\cal T}_\phi(\bk_1,\bk_2,\bk_3,\bk_4;{\bf K})
g_{\rm 21cm,l_1}(k_1) g_{\rm 21cm,l_2}(k_2) g_{\rm 21cm,l_3}(k_3) g_{\rm 21cm,l_4}(k_4) \nonumber \\
&&\times Y^*_{l_1m_1}(\hk_1) Y^*_{l_2m_2}(\hk_2) Y^*_{l_3m_3}(\hk_3) Y^*_{l_4m_4}(\hk_4)  \, .
\end{eqnarray}

The angular trispectrum associated with the $\delta^{(2)}$ term can be calculated numerically in a reasonable time, but the
exact numerical calculation of the angular trispectrum of 21-cm anisotropies associated with $\delta^{(3)}$ term is slow.
Here, we outline the analytical derivation of the trispectrum associated with the $\delta^{(2)}$ term, but for 
the $\delta^{(3)}$ term, following
an approach similar to prior calculations of the non-Gaussianity associated with the trispectrum of
gravitational evolution \cite{Scoc2}, we employ an approximation
with the angular averaged value $R_b\equiv\langle F_3(\bk_1,\bk_2,\bk_3)\rangle=682/189$ \cite{fry84} and ignore the
exact mode coupling resulting from the $F_3$ term. This assumption allows us to write $T_\delta^{(3)}=R_b[P(k_1)P(k_2)P(k_3)+..]$.

Compared to the derivation in Section~II where the trispectrum from primordial perturbations involve a coupling term which is
momentum-independent, the derivation of the trispectrum associated with non-linear evolution with a momentum-dependent
term is tedious. We take multipole moments of the $F_2$ term
\begin{equation}
F_2(\bk_1,\bk_2) = (4\pi) \sum_{l_a m_a} F_{2,l_a}(k_1,k_2) Y_{l_am_a}(\hk_1) Y^*_{l_am_a}(\hk_2)
\end{equation}
and since $F_2$ involves terms $(\bk_1 \cdot \bk_2)^n$ from $n=0,1,2$, $l_a$ takes the values of 0,1,2.
We outline the contribution for a
a specific combination of $(l_a,l_b)$ involving the expansion of the two $F_2$ terms,
the trispectrum calculation generally involves a term of the form
\begin{eqnarray}
&& \langle a_{l_1 m_1} a_{l_2 m_2} a_{l_3 m_3} a_{l_4 m_4} \rangle = 4 (4\pi)^6 (-i)^{\sum l_i}
\int \frac{d^3{\bf k}_1}{\left( 2\pi \right) ^3}...\int \frac{d^3{\bf k}_1}{\left( 2\pi \right) ^3} \int d^3K \int \frac{d^3{\bf r}_1}{\left( 2\pi \right) ^3} \int \frac{d^3{\bf r}_2}{\left( 2\pi \right) ^3} \\
&&\times \sum_{m_a m_b} Y_{l_am_a}(\hk_1) Y^*_{l_am_a}(\hat{\bf K}) Y_{l_bm_b}(\hk_3) Y^*_{l_bm_b}(\hat{\bf K}) F_{2,l_a}(k_1,K) F_{2,l_b}(k_3,K)P_\phi(K)P_\phi(k_1)P_\phi(k_3) \nonumber \\
&&\times e^{i {\bf r_1} \cdot (\bk_1+\bk_2+{\bf K})} e^{i {\bf r_2} \cdot (\bk_3+\bk_4-{\bf K})}
g_{\rm 21cm,l_1}(k_1) g_{\rm 21cm,l_2}(k_2) g_{\rm 21cm,l_3}(k_3) g_{\rm 21cm,l_4}(k_4)
Y^*_{l_1m_1}(\hk_1) Y^*_{l_2m_2}(\hk_2) Y^*_{l_3m_3}(\hk_3) Y^*_{l_4m_4}(\hk_4) \, , \nonumber
\end{eqnarray}
which can be simplified with Rayleigh expansion of the plane waves followed by angular integrals to arrive
after some tedious but straightforward algebra to
\begin{eqnarray}
&& \langle a_{l_1 m_1} a_{l_2 m_2} a_{l_3 m_3} a_{l_4 m_4} \rangle = \\
&& 4 (4\pi)^{12} (-i)^{l_1+l_3} \int k_1^ dk_1...\int k_4^2dk_2 \int K^2 dK \int r_1^2 dr_1 \int r_2^2 dr_2
F_{2,l_a}(k_1,K) F_{2,l_b}(k_3,K) P_\phi(K)P_\phi(k_1)P_\phi(k_3)  \nonumber \\
&&\times \sum_{m_am_b L_1 M_1 L_2 M_2 L_3 M_3 L_4} i^{\sum L_i} (-1)^{m_1+m_2+m_3+M_1+L_4}
g_{\rm 21cm,l_1}(k_1) g_{\rm 21cm,l_2}(k_2) g_{\rm 21cm,l_3}(k_3) g_{\rm 21cm,l_4}(k_4) \nonumber \\
&&\times j_{l_2}(k_2 r_1) j_{l_4}(k_4 r_2) j_{L_1}(k_1 r_1) j_{L_2}(k_3 r_2) j_{l_3}(K r_1) j_{L_4}(K r_2) h_{L_1l_1l_a} h_{L_2l_3l_b} h_{L_3l_2L_1}h_{L_4l_4L_2} \nonumber \\
&&\times \left(
\begin{array}{ccc}
L_1 & l_1 & l_a \\
M_1 & -m_1  & m_a
\end{array}
\right) \left(
\begin{array}{ccc}
L_2 & l_3 & l_b \\
M_2 & -m_3  &  m_b
\end{array}
\right) \left(
\begin{array}{ccc}
L_3 & l_2 & L_1 \\
M_3 & -m_2  & -M_1
\end{array}
\right) \left(
\begin{array}{ccc}
L_4 & l_4 & L_2 \\
M_4 & m_4  &  M_2
\end{array}
\right) \nonumber \\
&&\times \int d\hat{\bf K} Y^*_{l_am_a}(\hat{\bf K}) Y^*_{l_bm_b}(\hat{\bf K}) Y^*_{L_3 M_3}(\hat{\bf K}) Y_{L_4 M_4}(\hat{\bf K})  \nonumber
\end{eqnarray}

Including the exact form of the mode-coupling, we obtain the reduced angular
trispectrum of 21-cm anisotropies due to second-order gravitational perturbation evolution as
\begin{eqnarray}
&&{{\cal T}_{\rm grav}}^{l_1 l_2}_{l_3 l_4}(L) = 4 \left(\frac{2}{\pi}\right)^6 \int k_1^2 dk_1 ... \int k_4^2 dk_2 \int K^2 dK \int r_1^2 dr_1 \int r_2^2 dr_2 P_\phi(K)P_\phi(k_1)P_\phi(k_3)  \sum_{L_1 L_2 L_3 L_4} \\
&&\times g_{\rm 21cm,l_1}(k_1) g_{\rm 21cm,l_2}(k_2) g_{\rm 21cm,l_3}(k_3) g_{\rm 21cm,l_4}(k_4) j_{l_2}(k_2 r_1) j_{l_4}(k_4 r_2) j_{L_1}(k_1 r_1) j_{L_2}(k_3 r_2) j_{l_3}(K r_1) j_{L_4}(K r_2)  \nonumber \\
&&\times
\left[\frac{289}{441}S(0,0)+\frac{k_1k_3}{9K^2}S(1,1)+
\frac{136}{2205}S(0,2)+\frac{16}{11025}S(2,2)+\frac{17k_1}{63K}S(1,0)+\frac{17}{63K}S(0,1)+\frac{4k_1}{405K}S(1,2)+\frac{4k_3}{405K}S(2,1)\right]\,, \nonumber
\end{eqnarray}
where
\begin{eqnarray}
S(l_a,l_b) = \left\{
\begin{array}{ccc}
l_1 & l_2 & L \\
L_3 & l_a  & L_1
\end{array}
\right\} \left\{
\begin{array}{ccc}
l_3 & l_4 & L \\
L_4 & l_b  &  L_2
\end{array}
\right\}  h_{L_1l_1l_a} h_{L_2l_3l_b}
h_{L_3l_2L_1}h_{L_4l_4L_2} h_{L_4l_bL}h_{L_3l_aL} \, ,
\end{eqnarray}

As mentioned above, the angular trispectrum with $T_\delta^{(3)}$ term involves a calculation that is
numerically slow given the  mode coupling resulting from a term involving $F_3 \propto 1/(k_1+k_2+k_3)^2$ \footnote{
Analytically, one can still factor out the terms by making use of the Schwinger equality with
$(k_1+k_2+k_3)^{-2}=\int_0^\infty z e^{-z(k_1+k_2+k_3)}\, dz$}. Since we are considering the 3-1 angular power spectrum,
we can ignore the exact momentum dependence of the $\delta^{(3)}$ non-linear gravity and
take the angular averaged value of $F_3$. The resulting trispectrum in this case takes a simple form similar to that of
the primordial trispectrum with a momentum-independent coupling term, and we overestimate the covariance
between primordial and non-linear gravity trispectra to the 3-1 angular power spectrum estimator.
We do not reproduce the derivation of the angular trispectrum with $R_b$ given that it is similar to equation~(\ref{eqn:tng}).

\end{document}